%% file: plb_acp_kk_kkpi_pipi.tex
\documentclass{elsart}

\input epsf

\begin{document}

\setcounter{footnote}{0}

\begin{frontmatter}
\title{
Search for \emph{CP} violation in $D^0$ and $D^+$ decays
}
%
%
\input{author_list_plb.tex}
%
%
\begin{abstract}
 A high statistics sample of photoproduced charm particles from
the FOCUS (E831) experiment at Fermilab has been used to search for \emph{CP} 
violation in the Cabibbo suppressed decay modes 
$D^+ \to K^-K^+\pi^+$, $D^0 \to K^-K^+$ and $D^0 \to \pi^-\pi^+$. 
We have measured the following \emph{CP} asymmetry parameters:
$A_{CP}(K^-K^+\pi^+) = +0.006 \pm 0.011 \pm 0.005$, 
$A_{CP}(K^-K^+) = -0.001 \pm 0.022 \pm 0.015$ and 
$A_{CP}(\pi^-\pi^+) = +0.048 \pm 0.039 \pm 0.025$ where the first error 
is statistical and the second error is systematic. These 
asymmetries are consistent with zero with smaller errors than
previous measurements.
\end{abstract}

\end{frontmatter}

\section{Introduction}

 \emph{CP} violation occurs if the decay rate for a particle differs from the
decay rate of its \emph{CP}-conjugate particle \cite{BigiSanda}. \emph{CP} violation, 
which in the Standard Model (SM) is a consequence of a complex amplitude in the 
Cabibbo-Kobayashi-Maskawa (CKM) matrix, has been observed to date only 
in the neutral-kaon system. In charm meson decays (as well as in $K$ and $B$
decays), two classes of \emph{CP} violation exist: indirect and direct. 
In neutral-charm-meson decays, indirect \emph{CP} violation may arise 
due to $D^0-\overline{D}^0$ mixing. In the case of direct violation, 
\emph{CP} violating effects occur in a decay process only if the 
decay amplitude is the sum of two different parts, whose phases are made 
of a weak (CKM) and a strong contribution due to final state 
interactions (FSI) \cite{Buccella}. The weak contributions to the phases
change sign when going to the \emph{CP}-conjugate process, while the strong 
ones do not. In singly Cabibbo-suppressed $D$ decays, penguin terms 
in the effective Hamiltonian may provide the different phases of the 
two weak amplitudes.

 Compared to the strange and bottom sectors, the SM predictions of \emph{CP} 
violation for charm decays are much smaller \cite{Buccella,Bigi,Golden,Close}, 
making the charm sector a good place to test the SM and to 
look for evidence of new physics. In the SM, direct \emph{CP} 
violating asymmetries in $D$ decays are predicted to be largest in
singly Cabibbo-suppressed decays, at most $10^{-3}$, and non-existent in 
Cabibbo-favored and doubly Cabibbo-suppressed 
\hbox{decays \cite{BigiSanda}.} However, a \emph{CP} asymmetry could occur
in the decay modes $D \to K_s \rm{n}\pi$ due to interference
between Cabibbo-favored and doubly Cabibbo-suppressed decays.
 
 Experimentally, one looks at the Cabibbo-suppressed decay modes which have
the largest combination of branching fraction and detection efficiency. For 
this reason we select the all-charged decay modes $D^+ \to K^-K^+\pi^+$,
$D^0 \to K^-K^+$, and $D^0 \to \pi^-\pi^+$ (throughout this paper the charge 
conjugate state is implied, unless otherwise noted). 

 In $D$ decays the charged $D$ is self-tagging and the neutral $D$ is tagged
as either a $D^0$ or a $\overline{D}^0$ by using the sign of the bachelor 
pion in the $D^{*\pm}$ decay.

 Before searching for a \emph{CP} asymmetry we must account for
differences, at the production level, between $D$ and $\overline{D}$ 
in photoproduction (the hadronization process, in the presence
of remnant quarks from the nucleon, gives rise to production 
asymmetries \cite{Gardner} ). This is done by normalizing to the 
Cabibbo-favored modes $D^0 \to K^-\pi^+$ and $D^+ \to K^-\pi^+\pi^+$,
with the additional benefit that most of the corrections due to inefficiencies
cancel out, reducing systematic uncertainties. An implicit assumption is 
that there is no measurable \emph{CP} violation in the Cabibbo-favored decays.

The \emph{CP} asymmetry can be written as:

\begin{equation}
 A_{CP} = \frac{\eta(D)-\eta(\overline{D})}{\eta(D)+\eta(\overline{D})}  
\end{equation}
\noindent
 where $\eta$ is (considering for example the decay mode $D^0 \to K^-K^+$):
\begin{displaymath}
 \eta(D) = \frac{N(D^0 \rightarrow K^-K^+)}{N(D^0 \rightarrow K^-\pi^+)}
           \frac{\epsilon(D^0 \rightarrow K^-\pi^+)}{\epsilon(D^0 \rightarrow K^-K^+)}
\end{displaymath}
\noindent
where $N(D^0 \rightarrow K^-K^+)$ is the number of reconstructed candidate decays
and $\epsilon(D^0 \rightarrow K^-K^+)$ is the efficiency obtained from Monte 
Carlo simulations.
 
 The \emph{CP} asymmetry parameter measures the direct \emph{CP} 
asymmetry in the case of $D^+$ and the combined direct and indirect 
\emph{CP} asymmetries for $D^0$ \cite{Palmer}.

 The name FOCUS stands for {\bf Pho}toproduction of {\bf C}harm with an {\bf U}pgraded 
{\bf S}pectrometer with a lexical license. The word ``upgrade'' refers to the upgrade of 
the E687 (the predecessor experiment) spectrometer \cite{spectro}.

 Charmed particles were produced by the interaction of high energy photons, obtained by 
means of bremsstrahlung of electron and positron beams (with typically $300$ GeV endpoint energy),
with a beryllium oxide target. The mean energy of the photon beam was approximately $180$ GeV. 
The data were collected at Fermilab during the 1996--97 fixed-target run. More than 
$6.3 \times 10^9$ triggers were collected from which more than 1 million 
charmed particles have been reconstructed.
 
 The particles from the interaction are detected in a large-aperture magnetic spectrometer with 
excellent vertex measurement, particle identification and calorimetric capabilities. The vertex
detector consists of two systems of silicon microvertex detectors. The upstream system consists
of 4 planes interleaved with the experimental target, while the downstream system consists of 
12 planes of microstrips arranged in three views. These detectors provide high resolution separation 
of primary (production) and secondary (decay) vertices with an average proper time 
resolution of $30~\mathrm{fs}$ for 2-track vertices. 
The momentum of the charged particles is determined by measuring their deflections in two analysis 
magnets of opposite polarity with five stations of multiwire proportional chambers. Kaons and pions 
in the $D$-meson final states are well separated up to $60~{\rm GeV}/c$ of momentum using 
three multicell threshold \v{C}erenkov counters. 

\section{Selection criteria}

 The final states are selected using a candidate driven vertex algorithm \cite{spectro}.
A secondary vertex is formed from the reconstructed tracks and the 
momentum vector of the $D$ candidate is used as a {\it seed} to intersect the 
other tracks in the event to find the primary vertex. Once the 
production and decay vertices are determined, the distance $\ell$ between 
them and the relative error $\sigma_\ell$ are computed. Cuts on the
$\ell/\sigma_\ell$ ratio are applied to extract the $D$ signals from the 
prompt background. The topological configuration of the event is 
evaluated with four tests: the primary and secondary vertex confidence 
levels (minimum values of $1\%$ were required) and two measures of vertex 
isolation, a {\it no point-back isolation} and a {\it secondary vertex 
isolation}. The {\it no point-back isolation} cut requires that the maximum confidence level 
for a candidate-$D$ daughter track to form a vertex with the tracks from the primary vertex be
less than a certain threshold. The {\it secondary vertex isolation} cut requires that the maximum 
confidence level for another track to form a vertex with the $D$ candidate be less than 
a certain threshold. The analyses differ mainly in the way the particle identification is handled 
and, less importantly, in the way the vertex cuts are applied. To minimize 
the systematic error we use identical vertex cuts on the signal and normalizing
modes. 

\begin{figure}[ht!]
\vspace{13cm}
\epsfbox{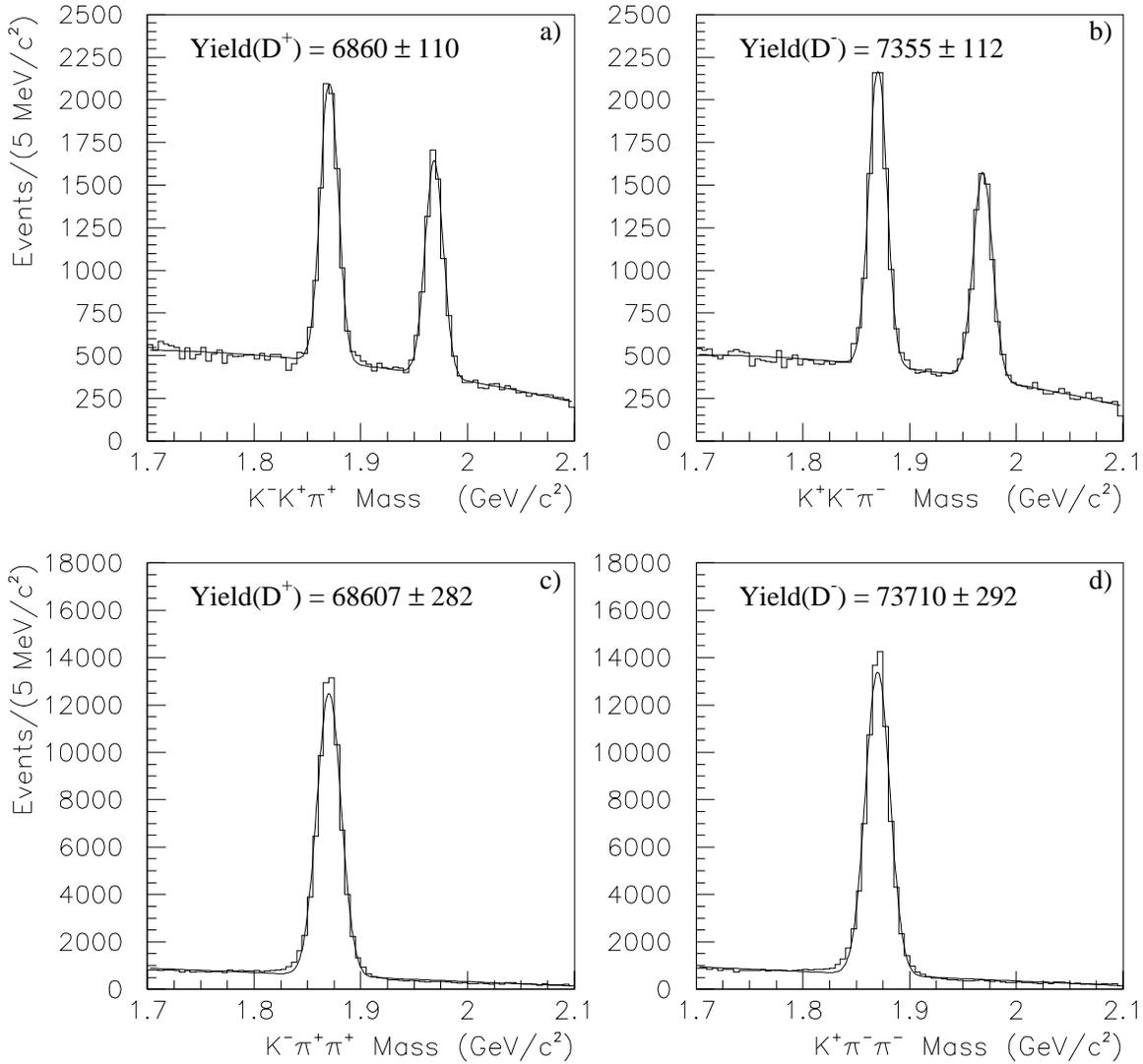} 
\caption{(a) $K^-K^+\pi^+$ invariant mass distribution, 
 (b) $K^+K^-\pi^-$ invariant mass distribution, (c) $K^-\pi^+\pi^+$
 invariant mass distribution, (d) $K^+\pi^-\pi^-$ invariant mass distribution.
 The fits (solid curves) are described in the text and the numbers quoted 
 are the yields.}
\end{figure}

 In the $D^+ \to K^-K^+\pi^+$ analysis, we require $\ell/\sigma_\ell > 10$, the {\it no point-back isolation}
must be less than $20\%$, and the {\it secondary vertex isolation} less than $0.1\%$. The 
vertices (primary and secondary) have to lie inside a {\it fiducial volume}\footnote{The reason 
for this cut lies in the presence of a trigger counter just upstream of the second microstrip device,
therefore we defined the {\it fiducial volume} as the target region between the first slab of the 
experimental target and this trigger counter.}, the $D$ momentum must be in the range 
$25 < P < 250~{\rm GeV}/c$ (a very loose cut) and the primary vertex must be formed with 
at least two reconstructed tracks, in addition to the {\it seed} track. 
The \v{C}erenkov particle identification cuts used in FOCUS are based on likelihood ratios 
between the various stable particle identification hypotheses. These likelihoods are computed 
for a given track from the observed firing response (on or off) of all cells within the 
track's ($\beta =1$) \v{C}erenkov cone for each of our three \v{C}erenkov counters. The 
product of all firing probabilities for all cells within the three \v{C}erenkov cones produces 
a $\chi^2$-like variable 
$W_i = -2 \ln (\mathrm{Likelihood})$ where $i$ ranges over the electron, pion, kaon and proton 
hypotheses (see reference \cite{jim} for more details). We require $W_\pi - W_K > 1$ for the
tracks reconstructed as a kaon, and a pion consistency cut for the pion tracks.
The pion consistency cut requires that no particle hypothesis is favored over the pion hypothesis 
with a $\Delta W = W_\pi - W_{min}$ exceeding 2. To remove contamination from the  
$D_s^+ \to K^-K^+\pi^+$ signal due to \v{C}erenkov misidentified background from the 
decay mode $D^+ \to K^-\pi^+\pi^+$, we employ an {\it anti-reflection} cut which rejects 
candidates which, when reconstructed as $K^-\pi^+\pi^+$, are consistent with the 
$D^+$ hypothesis (we also reject events whose $K^-K^+$ mass exceeds $1.84$ GeV/$c^2$ in order 
to exclude background due to $D^{*+} \to D^0\pi^+ \to (K^-K^+)\pi^+$). This cut has no effect 
in the vicinity of the $D^+ \to K^-K^+\pi^+$ signal peak.

 In Fig.1, the invariant mass plots obtained with this set of cuts 
for the decay modes 
$D^+ \to K^-K^+\pi^+$, $D^- \to K^+K^-\pi^-$, and the normalizing decays 
$D^+ \to K^-\pi^+\pi^+$ and $D^- \to K^+\pi^-\pi^-$ are shown. In the 
$D^+ \to K^-\pi^+\pi^+$ analysis there is an additional cut on the $D^0$ mass formed
by a kaon and a pion to remove the $D^{*+} \to D^0\pi^+ \to (K^-\pi^+)\pi^+$ decay chain. 
The $KK\pi$ invariant mass distributions are fit with a Gaussian for the $D^+$ signal, a 
second Gaussian for the $D^+_s$ signal, and a quadratic polynomial for the background. 
A binned maximum likelihood fit finds $6860 \pm 110 ~ D^+ \to K^-K^+\pi^+$ and 
$7355 \pm 112 ~ D^- \to K^+K^-\pi^-$ events. The fit for the normalizing modes
(fit with a Gaussian plus a linear polynomial) gives 
$68607 \pm 282 ~ D^+ \to K^-\pi^+\pi^+$ and $73710 \pm 292 ~ D^- \to K^+\pi^-\pi^-$
events.

 In the $D^0 \to K^-K^+$ analysis, the sign of the bachelor 
pion in the $D^{*\pm}$ decay chain $D^{*+(-)} \to D^0(\overline{D}^0)\pi^{+(-)}$ is used to 
identify the neutral $D$ as either a $D^0$ or a $\overline{D}^0$. We require that the mass 
difference between the $D^0$ and the $D^*$ mass be within $4$ MeV/$c^2$ with respect to the nominal
mass difference \cite{pdg}. We use $\ell/\sigma_\ell > 8$, while the {\it no point-back isolation} and
the {\it secondary vertex isolation} cuts are unnecessary because the $D^*$ tag sufficiently
reduces the background. All the other cuts, except the {\it anti-reflection} cuts, 
are the same as those used in the $D^+ \to K^-K^+\pi^+$ analysis.

\begin{figure}[ht!]
\vspace{13cm}
\epsfbox{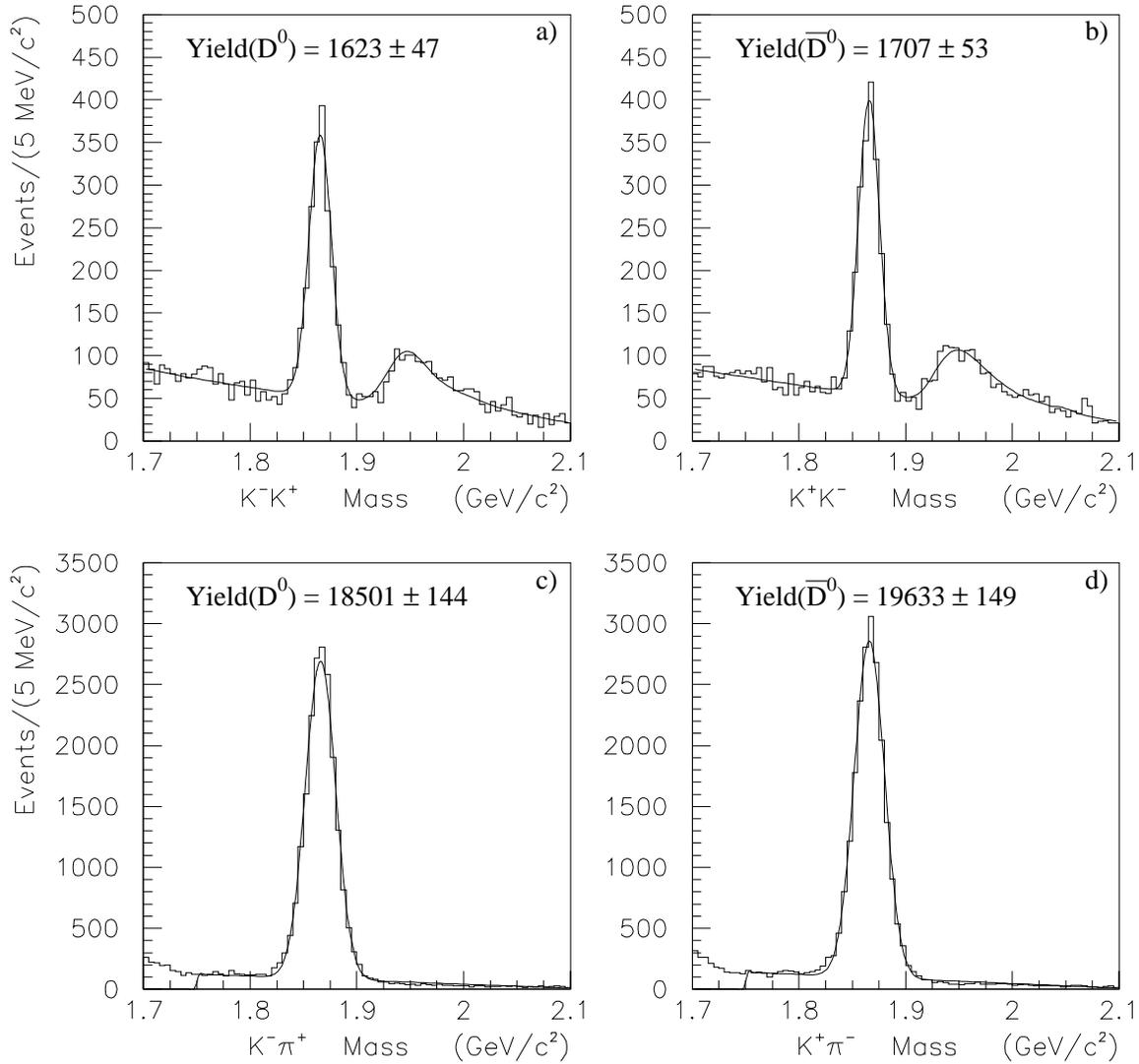} 
\caption{ (a) $K^-K^+$ invariant mass distribution, 
 (b) $K^+K^-$ invariant mass distribution, (c) $K^-\pi^+$
 invariant mass distribution, (d) $K^+\pi^-$ invariant mass distribution.
 The fits (solid curves) are described in the text and the numbers quoted 
 are the yields.}
\end{figure}

 In Fig.2, we show the invariant mass plots, obtained with this set of cuts, for the 
decay modes $D^0 \to K^-K^+$, $\overline{D}^0 \to K^+K^-$, and the normalizing decays 
$D^0 \to K^-\pi^+$ and $\overline{D}^0 \to K^+\pi^-$. The peak in the $KK$ invariant mass plots
at $\approx$ 1.95 GeV/$c^2$
is due to the reflection of the $D^0 \rightarrow K^-\pi^+$ mode when the pion is misidentified 
as a kaon. A Monte Carlo simulation of this reflection reproduces the shape
observed in the data. Consequently, the $K^-K^+$ invariant mass distributions are fit with 
a Gaussian for the $D^0$ signal, a function obtained by smoothing the reflection 
peak (only the shape of this reflection peak is modeled by our Monte Carlo 
simulation, the amplitude of this peak is given by a free parameter of the fit) 
and a quadratic polynomial for the background. From a binned maximum likelihood fit 
we find $1623 \pm 47 ~ D^0 \to K^-K^+$ and $1707 \pm 53 ~ \overline{D}^0 \to K^+K^-$ events. 
The fit for the normalizing modes (fit with a Gaussian plus a linear polynomial 
and excluding the low mass region to avoid possible contamination due to other charm 
hadronic decays involving an additional $\pi^0$ ) gives $18501 \pm 144 ~ D^0 \to K^-\pi^+$ 
and $19633 \pm 149 ~ \overline{D}^0 \to K^+\pi^-$ events.

\begin{figure}[ht!]
\vspace{5.5cm}
\epsfbox{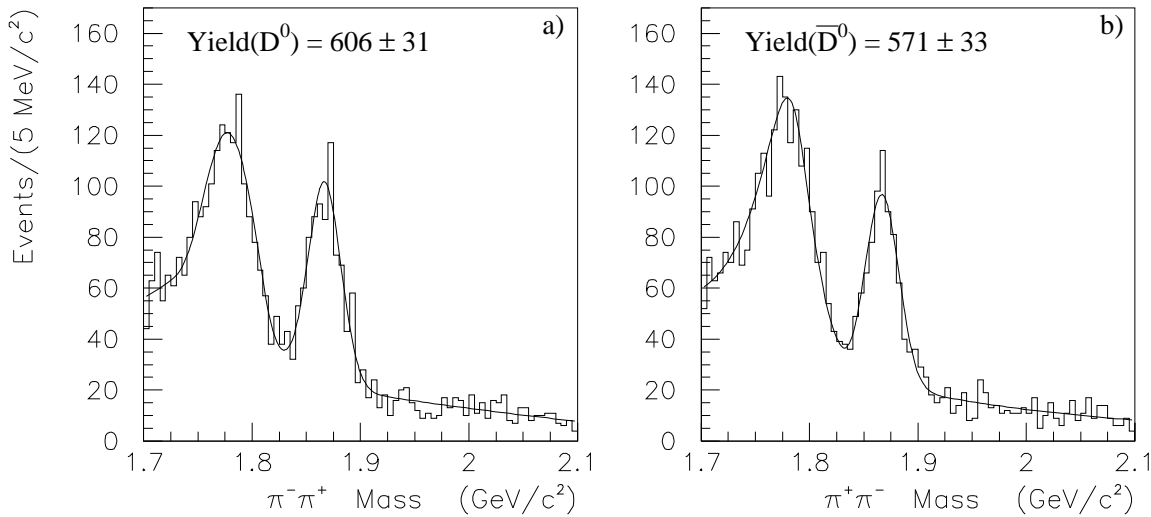} 
\caption{(a) $\pi^-\pi^+$ invariant mass distribution, 
 (b) $\pi^+\pi^-$ invariant mass distribution. 
 The fits (solid curves) are described in the text and the numbers quoted 
 are the yields. }
\end{figure} 

  The $D^0 \to \pi^-\pi^+$ analysis is identical to the $D^0 \to K^-K^+$ analysis with 
the exception of the \v{C}erenkov identification. In order to reduce the large reflection 
peak from the $D^0 \rightarrow K^-\pi^+$ mode, a tight \v{C}erenkov identification
requirement ($W_K - W_\pi > 1$) for the pion tracks is implemented.

 In Fig.3 the invariant mass plots for the decay modes $D^0 \to \pi^-\pi^+$ 
and $\overline{D}^0 \to \pi^+\pi^-$ are shown. As in the $D^0 \to K^-K^+$ 
case, the peak at $\approx 1.75 \, {\rm GeV}/c^2$ is due to the reflection of the 
$D^0 \rightarrow K^-\pi^+$ mode. Again our Monte Carlo simulation reproduces the shape 
observed in the data. The $\pi^-\pi^+$ invariant mass distributions are fit with a Gaussian
for the $D^0$ signal, a function obtained by smoothing the reflection peak, and a quadratic 
polynomial for the background. A binned maximum likelihood fit gives 
$606 \pm 31 ~ D^0 \to \pi^-\pi^+$ and $571 \pm 33 ~ \overline{D}^0 \to \pi^+\pi^-$ events.

\section{\emph{CP} asymmetry measurements}

 Because the efficiency is strongly dependent on the $D$ momentum, it is necessary to verify that
the observed momentum spectrum is reproduced by the Monte Carlo simulation. A mismatch could 
generate a false asymmetry. Fig.4 shows the $D^0$ momentum for the decay mode 
$D^0 \to K^-\pi^+$ for real data (points with errors) and Monte Carlo data (histogram). The 
$D^0$ momentum spectrum is obtained (both in real data and Monte Carlo data) by subtracting 
the events contained in the sideband regions from the events in the signal region. The two 
histograms are normalized by area. The agreement is good.

\begin{figure}[ht!]
\vspace{5.5cm}
\epsfbox{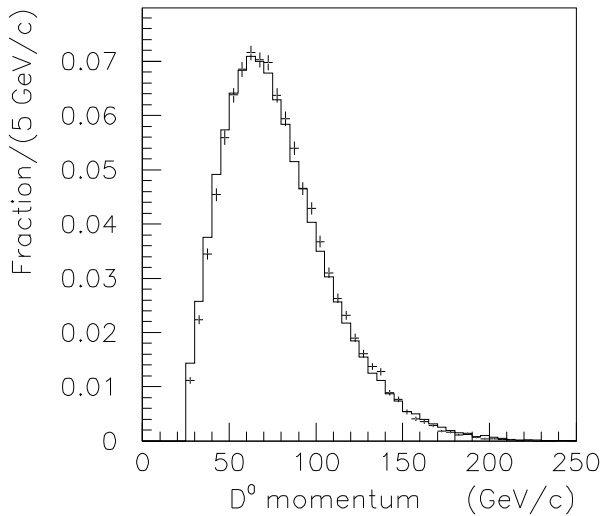}
\caption{$D^0$ momentum for the decay mode $D^0 \to K^-\pi^+$. The points with errors 
 correspond to the real data, the histogram to the Monte Carlo data.}
\end{figure}

 The measured asymmetries are reported in table 1 along with a comparison to previous measurements.
 The analysis of the decay mode $D^+ \to K^-K^+\pi^+$ is complicated by the possibility of 
intermediate resonant states, such as $K^{*0}K^+$ and $\phi\pi^+$. We think that in this case a 
Dalitz plot analysis is the appropriate tool to investigate \emph{CP} asymmmetry effects. 
This topic will be the subject of a future paper.
   
 Our asymmetry measurements have been tested by modifying each of the vertex 
cuts individually; the results are always consistent.

\begin{table}[hb!]
\caption{Measured \emph{CP} asymmetries ($\times 10^{-2}$).}
\begin{tabular}{|l|lll|}
\hline
 Experiment                & $D^+ \to K^-K^+\pi^+$ & $D^0 \to K^-K^+$    &  $D^0 \to \pi^-\pi^+$ \\
\hline
  E687 \cite{E687}          & $-3.1 \pm 6.8$ & $+2.4 \pm 8.4$ &  	                      \\
  CLEO II \cite{CLEO}       &                & $+8.0 \pm 6.1$ &  	                      \\
  E791 \cite{E791_+,E791_0} & $-1.4 \pm 2.9$ & $-1.0 \pm 4.9 \pm 1.2$ & $-4.9 \pm 7.8 \pm 3.0$ \\
  This measurement & $+0.6 \pm 1.1 \pm 0.5$ & $-0.1 \pm 2.2 \pm 1.5$         & $+4.8 \pm 3.9 \pm 2.5$ \\
\hline  
\end{tabular}
\end{table}

 The systematic errors on these measurements reflect uncertainties
in reconstruction efficiency, \v{C}erenkov particle identification, and 
hadronic absorption of secondaries in the target and spectrometer
materials. The estimates were obtained by splitting our data 
into independent samples depending on $D$ momentum and 
the different periods in which the data were collected. The main reason for this
time depedence is the insertion of the upstream silicon system (in the target region) during 
the 1997 fixed-target run period. A technique modeled after the {\it S-factor method} from the 
Particle Data Group \cite{pdg} was used to separate true systematic variations from statistical 
fluctuations. The asymmetry is evaluated for each of the statistically independent subsamples and a 
{\it scaled variance} is calculated; the {\it split sample} variance is defined as the difference
between the reported statistical variance and the scaled variance if the scaled variance 
exceeds the statistical variance. For all decays studied, the scaled variance is less than
the statistical variance, so it does not contribute to the systematic uncertainty. 
The evaluation of systematic effects related to different fit 
procedures is performed on the whole sample. The asymmetry is calculated using various fit 
conditions (these include the choice of the estimator, the background shape and,
in the case of the $D^0$ analysis, the shape of the reflection peak) and the sample variance is
used because the fit variants are all {\it a priori} likely. To obtain the final systematic error,
the variance from the different fitting procedures and a further contribution, due to the 
uncertainties in the efficiency calculation and finite statistics of the Monte Carlo events,
are then added in quadrature. Table 2 shows the contribution of each of these sources to the 
total systematic uncertainty for the \emph{CP} asymmetry measurements.

 The measured asymmetries are consistent with zero within the errors. 

\begin{table}
\caption{Contributions to the systematic uncertainty for the measured asymmetries.}
\begin{tabular}{|l|lll|}
\hline
 Source & $A_{CP}(K^-K^+\pi^+)$ & $A_{CP}(K^-K^+)$ & $A_{CP}(\pi^-\pi^+)$ \\
\hline
 Split sample           & {\rm no contribution} & {\rm no contribution} & {\rm no contribution} \\
 Fit variant            & 0.0016                &  0.0144               &  0.0243               \\
 Monte Carlo statistics & 0.0050                &  0.0054               &  0.0055               \\
\hline
 Total systematic error & 0.0053                &  0.0154               &  0.0249               \\
\hline 
\end{tabular}
\end{table}
 
\section{Summary and Conclusions}

 We have searched for \emph{CP} violation in the Cabibbo-suppressed decay modes 
$D^+ \to K^-K^+\pi^+$, $D^0 \to K^-K^+$ and $D^0 \to \pi^-\pi^+$ using a high 
statistics sample of photoproduced charm particles from
the FOCUS (E831) experiment at Fermilab. We have measured the following 
\emph{CP} asymmetry parameters: $A_{CP}(K^-K^+\pi^+) = +0.006 \pm 0.011 \pm 0.005$, 
$A_{CP}(K^-K^+) = -0.001 \pm 0.022 \pm 0.015$ and 
$A_{CP}(\pi^-\pi^+) = +0.048 \pm 0.039 \pm 0.025$ where the first error is statistical
and the second error is systematic. 

These asymmetries are consistent with zero and represent a substantial 
improvement over previous measurements.

\vspace{1.cm}

We wish to acknowledge the assistance of the staffs of Fermi National
Accelerator Laboratory, the INFN of Italy, and the physics departments of
the
collaborating institutions. This research was supported in part by the U.~S.
National Science Foundation, the U.~S. Department of Energy, the Italian
Istituto Nazionale di Fisica Nucleare and Ministero dell'Universit\`a e
della
Ricerca Scientifica e Tecnologica, the Brazilian Conselho Nacional de
Desenvolvimento Cient\'{\i}fico e Tecnol\'ogico, CONACyT-M\'exico, the
Korean
Ministry of Education, and the Korean Science and Engineering Foundation.
 
%
%
%
%
%

%
\end{document}

%% file: author_list_plb.tex
The FOCUS Collaboration

\author[davis]{J.M.~Link},
\author[davis]{V.S.~Paolone\thanksref{atpitt}},
\author[davis]{M.~Reyes\thanksref{atmex}},
\author[davis]{P.M.~Yager},
\author[cbpf]{J.C.~Anjos},
\author[cbpf]{I.~Bediaga},
\author[cbpf]{C.~G\"obel\thanksref{aturug}},
\author[cbpf]{J.~Magnin\thanksref{atbogota}},
\author[cbpf]{J.M.~de~Miranda},
\author[cbpf]{I.M.~Pepe\thanksref{atbahia}},
% ...

\address[davis]{University of California, Davis, CA 95616}
\address[cbpf]{Centro Brasileiro de Pesquisas F\'{\i}sicas, Rio de Janeiro,
RJ, Brazil}
\address[cine]{CINVESTAV, 07000 M\'exico City, DF, Mexico}
\address[cu]{University of Colorado, Boulder, CO 80309}
\address[fnal]{Fermi National Accelerator Laboratory, Batavia, IL 60510}
\address[fras]{Laboratori  Nazionali di Frascati dell'INFN, Frascati, Italy,
      I-00044}
\address[ui]{University of Illinois, Urbana-Champaign, IL 61801}
\address[iu]{Indiana University, Bloomington, IN 47405}
\address[koru]{Korea University, Seoul, Korea 136-701}
\address[milan]{INFN and University of Milano, Milano, Italy}
\address[anc]{University of North Carolina, Asheville, NC 28804}
\address[pavia]{Dipartimento di Fisica Nucleare e Teorica and INFN,
Pavia, Italy}
\address[pr]{University of Puerto Rico, Mayaguez, PR 00681}
\address[sc]{University of South Carolina, Columbia, SC 29208}
\address[ut]{University of Tennessee, Knoxville, TN 37996}
\address[vandy]{Vanderbilt University, Nashville, TN 37235}
\address[wisc]{University of Wisconsin, Madison, WI 53706}
\address[yonsei]{Yonsei University, Seoul, Korea 120-749}

\thanks[atpitt]{Present Address: University of Pittsburgh, Pittsburgh,
PA 15260}
\thanks[atmex]{Present Address: Instituto de Fisica y
Matematicas, Universidad Michoacana de
San Nicolas de Hidalgo, Morelia, Mich., Mexico 58040}
\thanks[aturug]{Present Address: Instituto de F\'isica, Faculdad de
Ingenier\'i a, Univ. de la Rep\'ublica, Montevideo, Uruguay}
\thanks[atbogota]{Present Address: Universidad de los Andes, Bogota,
Colombia}
\thanks[atbahia]{Present Address: Instituto de F\'isica, Universidade
Federal da Bahia, Salvador, Brazil}
\thanks[atmilan]{Present Address: INFN sezione di Milano, Milano,
Italy}
\thanks[atpr]{Present Address: University of Puerto Rico, Mayaguez,
PR  00681}
\thanks[atncar]{Present Address: National Center for Atmospheric Research, Boulder, CO} 
\thanks[atlbl]{Present Address: Lawrence Berkeley Lab, Berkeley, CA
94720}
\thanks[atbrescia]{Present Address:
Dipartimento di Chimica e Fisica per l'Ingegneria e per i Materiali,
Universit\`a di Brescia  and  INFN sezione di Pavia}
\thanks[atromania]{Present Address: Nat. Inst. of Phys. and Nucl. Eng., Bucharest, Romania}
\thanks[atcu]{Present Address: University of Colorado, Boulder, CO 80309}
\thanks[atlucent]{Present Address: Lucent Technology}
\thanks[atagusta]{Present Address: Augusta Technical Inst., Augusta, GA
30906}
\thanks[atadams]{Present Address: Adams State College, Alamosa, CO 81102} 

%% file: plb_acp_kk_kkpi_pipi.bbl
\begin{thebibliography}{999}
\bibitem{BigiSanda}A general reference textbook for \emph{CP}  violation is:\\
 I.I.~Bigi and A.I.~Sanda, {\em CP Violation\/}, Cambridge University Press (2000).
\bibitem{Buccella} F.~Buccella {\it et al.}, {\em Phys.Rev.\/} {\bf D51} (1995) 3478.
\bibitem{Bigi} I.I.~Bigi, {\em Proceedings of the Tau-Charm Factory Workshop\/}, 
  (SLAC 1989), SLAC-Report-343.
\bibitem{Golden} M.~Golden and B.~Grinstein, {\em Phys.~Lett.\/} {\bf B222} (1989) 501.
\bibitem{Close} F.~Close and H.~Lipkin, {\em Phys.~Lett.\/} {\bf B372} (1996) 306.
\bibitem{Gardner} E687 Collaboration, P.L.~Frabetti {\it et al.}, {\em Phys.~Lett.\/} 
  {\bf B370} (1996) 222.
\bibitem{Palmer} W.F.~Palmer and Y.L.~Wu, {\em Phys.~Lett.\/} {\bf B350} (1995) 245.
\bibitem{spectro} E687 Collaboration, P.L.~Frabetti {\it et al.}, {\em Nucl.~Instr.~Meth.\/}
  {\bf A320} (1992) 519.
\bibitem{jim} FOCUS Collaboration, J.M.~Link {\it et al.}, FERMILAB-Pub-00/091-E and
  \hbox{hep-ex/0004034}, submitted to {\em Phys.~Lett.\/} {\bf B}.
\bibitem{pdg} Particle Data Group, C.~Caso {\it et al.}, {\em Eur.~Phys.~J.\/} {\bf C3} 
  (1998) 1.
\bibitem{E687} E687 Collaboration, P.L.~Frabetti {\it et al.}, {\em Phys.~Rev.\/} {\bf D50} 
  (1994) R2953.
\bibitem{CLEO} CLEO Collaboration, J.~Bartelt {\it et al.}, {\em Phys.~Rev.\/} {\bf D52}
  (1995) 4860.
\bibitem{E791_+} E791 Collaboration, E.M.~Aitala {\it et al.}, {\em Phys.~Lett.\/} {\bf B403}
  (1997) 377.  
\bibitem{E791_0} E791 Collaboration, E.M.~Aitala {\it et al.}, {\em Phys.~Lett.\/} {\bf B421}
  (1998) 405.
\end{thebibliography}
